# In-situ real-time observation of photo-induced nanoscale azo-polymer motions using high-speed atomic force microscopy combined with an inverted optical microscope

Keishi Yang[a], Feng-Yueh Chan[b], Hiroki Watanabe[c], Shingo Yoshioka[a], Yasushi Inouye[ad], Takayuki Uchihashi[bc], Hidekazu Ishitobi[ad], Prabhat Verma[a], and Takayuki Umakoshi*[ae]

a Department of Applied Physics, Osaka University, 2-1 Yamadaoka, Suita, Osaka 565-0871, Japan

b Department of Physics, Nagoya University, Aichi 464-8602, Japan

c Exploratory Research Center on Life and Living Systems (ExCELLS), National Institutes of Natural Sciences, Higashiyama 5-1, Myodaiji, Okazaki, 444–0864 Japan

d Graduate School of Frontier Biosciences, Osaka University, 1-3 Yamadaoka, Suita, Osaka 565-0871, Japan

e Institute of Advanced Co-Creation Studies, Osaka University, 2-1 Yamadaoka, Suita, Osaka 565-0871, Japan

*E-mail: umakoshi@ap.eng.osaka-u.ac.jp

**ABSTRUCT**

High-speed atomic force microscopy (HS-AFM) is an indispensable technique in the biological field owing to its excellent imaging capability for the real-time observation of biomolecules with high spatial resolution. Furthermore, recent developments have established a tip-scan stand-alone HS-AFM that can be combined with an optical microscope, drastically improving its versatility for studying various complex phenomena. Although HS-AFM has mainly been used in biology, it has considerable potential to contribute to various research fields. One of the great candidates is a photoactive material, such as an azo-polymer, which plays a vital role in multiple optical applications because of its unique nanoscale motion under light irradiation. In this study, we demonstrate the *in-situ* real-time observation of nanoscale azo-polymer motion by combining tip-scan HS-AFM with an optical system, allowing HS-AFM observations precisely aligned with a tightly focused laser position. We successfully observed the dynamic evolution of unique morphologies in azo-polymer films, attributed to photoinduced nano-movements. Moreover, real-time topographic line profile analyses facilitated precise and quantitative investigations of morphological changes, which provided novel insights into the deformation mechanism. This significant demonstration would pave the way for the application of HS-AFM in wide research fields, from biology to material science and physical chemistry.

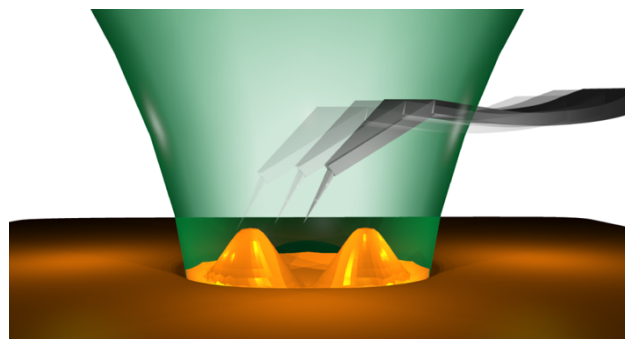

**INTRODUCTION**

High-speed atomic force microscopy (HS-AFM) is a well-recognized tool for capturing biological dynamic motions, including conformational changes in proteins with high spatiotemporal resolutions, (1) enabled via several technical developments such as small cantilevers, fast scanners, and dynamic feedback control. It has contributed significantly to numerous biological discoveries by providing movies of molecular dynamics at the single-molecule level, including walking myosin V, (2) rotary motion of rotorless $F_1$-ATPs, (3) DNA cleavage by CRISPR-Cas9 (4) and so on. (5-11) It has already become an indispensable technique in the field of biology.

In addition, recent technical developments in HS-AFM have even improved its analytical capability, which has expanded its possibilities for widespread use in various fields. (12-15) Among several technical improvements, tip-scan-type HS-AFM is of great importance as a critical development. (16,17) In the original HS-AFM, the tip is fixed while the sample stage is scanned rapidly. In contrast, the tip-scan HS-AFM is a stand-alone system for scanning the tip, which can be mounted on an inverted optical microscope. (17) It also enabled the implementation of complex sample manipulation, such as uniaxial stretching. (15) It was originally aimed at biological applications for studying complex phenomena by combining HS-AFM with optical microscopes. (16-18) However, the superior capability of the tip-scan HS-AFM can contribute to various other fields, such as material science and physical chemistry. It holds great promise as a technique for providing powerful and unique measurements by incorporating it with various optical techniques. For example, by irradiating a sample with a tightly focused laser and precisely adjusting the tip position to the focused laser spot, the dynamic processes locally induced in novel photoactive materials can be observed at the nanoscale in real time.

As novel photoactive materials, azo-polymers are of great importance in material science. Azo-polymers are advanced organic materials containing azobenzene molecules that induce photoisomerization reactions between the *trans-* and *cis-* forms under light irradiation, as shown in the inset of Fig. 1. When made into thin films, they exhibit a unique morphology on their surface because of the polymer movements caused by photoisomerization reactions. (19-26) Therefore, they have been considered promising materials for various optical applications such as optical data storage, grating-based devices, and photomechanical actuators. (27-29) However, the detailed mechanisms of the deformation process of azo-polymer films, which are crucial for fully utilizing the unique properties of azo-polymers for actual applications, remain elusive. In previous studies, conventional AFM has been used to study the azo-polymer deformation process. (21-26) To understand the temporal changes in the surface morphologies, distinct deformed surfaces were created at varying laser exposure durations. Subsequent to laser irradiation, AFM images of these pre-deformed structures were captured as time-lapse sequences to indirectly investigate temporal evolution. However, this indirect method provides limited information. Unfortunately, it is also technically difficult to analyze a process with high temporal resolution. The *in-situ* observation of azo-polymer movement would provide more reliable insights to elucidate the deformation mechanism.

In this study, we applied tip-scan HS-AFM for recording a real-time movie of the photoinduced nanoscale movements of an azo-polymer film. Compared with conventional AFM, HS-AFM enables direct *in-situ* observation of azo-polymer deformation with high temporal resolution, which allows visualization of the formation of individual structures in real time. To this end, we constructed a tip-scan HS-AFM combined with an optical setup. In the optical setup, the laser focus position was precisely adjustable to the tip position to perform HS-AFM imaging, while the photoactivated azo-polymer motions were induced by a tightly focused laser. We successfully demonstrated *in-situ* HS-AFM observations of azo-polymer movements with high spatiotemporal resolution. We also confirmed that the azo-polymers formed different surface morphologies depending on the polarization direction of the incident light. Furthermore, we found that the morphologies were slightly different from each other even when they were created under the same experimental conditions, highlighting the importance of observing individual structures using HS-AFM, unlike the assembly of AFM images of several different structures. These results indicate that tip-scan HS-AFM is highly beneficial for elucidating the mechanism of azo-polymer deformation. We also believe that tip-scan HS-AFM can be a significant tool for researchers to utilize it in their own research fields, from material science to many other fields.

**EXPERIMENTAL**

**Optical setup combined with HS-AFM**

To observe azo-polymer movements, a tip-scan HS-AFM was custom-built (Fig. 1), based on a previous study. (19) A miniaturized microcantilever (Olympus, BL-AC10DS-A2), 9 μm in length, 2 μm in width, and 100 nm in thickness, was used for high-speed imaging. The resonance frequency of this cantilever in air is approximately 1500 kHz. The stand-alone tip-scan HS-AFM was mounted on an inverted optical microscope (Nikon, ECLIPSE Ti2). The incident laser beam (Cobolt, 0532-04-01-0100-700, λ:532 nm), expanded using a beam expander, was focused from the bottom side of the samples through an oil-immersion objective lens (Olympus, NA1.45, ×100) of the inverted optical microscope. The opposite surfaces of the samples were probed by HS-AFM with the tip set at the top, as shown in the inset of Fig. 1. The power of the incident light was adjusted to approximately 4 nW at the sample plane using ND filters. The focal size was approximately 220 nm due to the diffraction limit. Waveplates and polarizers were used to control the polarization of the incident light. To observe the surface deformation process using HS-AFM, the position of the focus spot was precisely adjusted to the position of the tip. For this, home-made piezo-mirror scanners were installed to precisely control the position of the focal spot in the lateral direction. Another piezo scanner was installed on the sample stage to change the position of the sample. A mechanical shutter (Thorlabs, SHB05) was used to control the laser exposure time. All HS-AFM measurements were conducted in air.

**Preparation of azo-polymer film**

An azo-polymer film was prepared from poly[4′-[[2-(methacryloxy)ethyl]ethyl]amino-4-nitroazobenzene] (PMA-DR1). PMA-DR1 (Sigma-Aldrich, 579009) was dissolved in chloroform and stirred for 24 h at room temperature. The PMA-DR1 solution was filtered twice through a membrane filter with a pore size of 200 nm. The solution was spin-coated on a cover slip at 300 rpm for 3 s, and subsequently at 1,500 rpm for 60 s. The spin-coated film was heated in an oven for 1 h at 110 °C to remove the remaining solvent. The thickness of the azo-polymer film was approximately 30 nm.

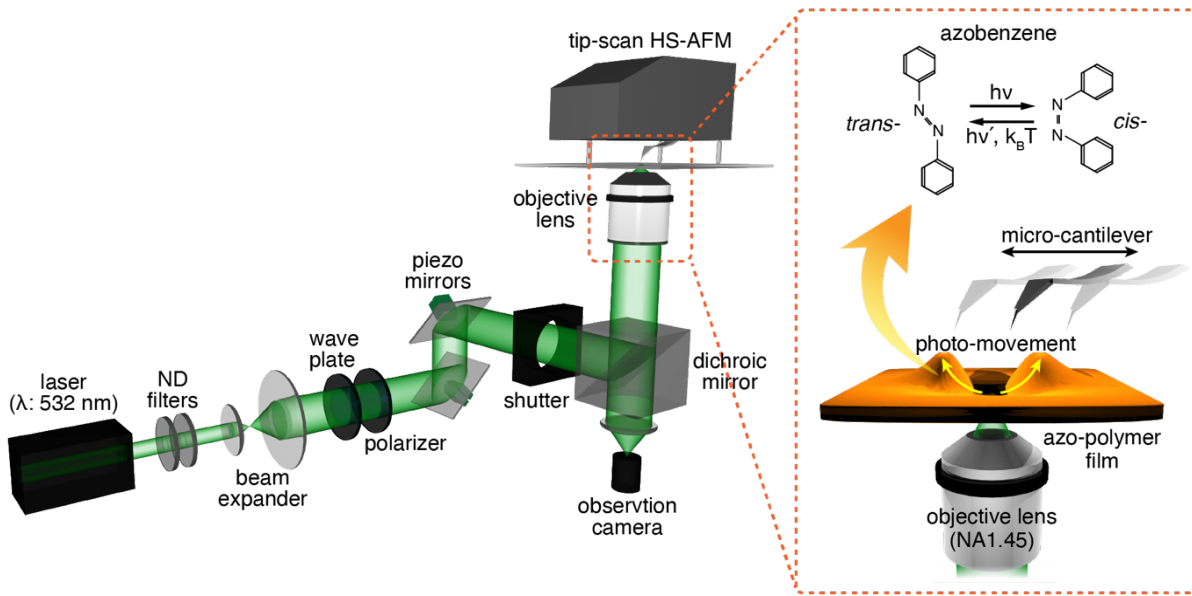

**Figure 1.** (a) Schematic of the experimental setup for the *in-situ* HS-AFM observation of azo-polymer deformation.

**RESULTS**

**HS-AFM observation of azo-polymer deformation**

Using the constructed setup, azo-polymer deformation was successfully observed in real time at an imaging rate of 2 frames per second (fps), as shown in Fig. 2(a) (Movie S1). Linearly polarized light was used in this experiment; the polarization direction is indicated by the green arrow in Fig. 2(a). The focus position and size are indicated by the green circle. Upon initiating light irradiation at 0 s by opening the mechanical shutter, the sample surface, which was initially flat, began to display two lobes aligned with the polarization direction of the incident light. A pit also formed between the two lobes. This unique pattern is similar to that reported in a previous study, (22) where conventional AFM was used after a complete exposure. This confirmed that, in our measurements, where we continuously measure AFM images during the exposure, azo-polymer movements can be observed with high temporal resolution using the HS-AFM technique. Over time, the two lobes became higher and the pit became deeper, with the left lobe being slightly higher than the right lobe. In addition, the distance between the lobes increased. This change in the surface morphology during the exposure, which is observed for the first time, is attributed to the azo-polymer moving outward from the center of the focus spot along the polarization direction of the incident light, as previously reported. (22) With linearly polarized light irradiation, the azobenzene molecules are oriented in the direction perpendicular to the incident light polarization during *trans* ↔ *cis* photoisomerization. (30) According to the anisotropic nature induced by the molecular orientation, anisotropic photo-fluidic force is generated under the condition that the polymer is softened by the repeated photoisomerization reactions between the *trans-* and *cis-* forms. (31) This anisotropic photo-fluidic force induces the formation of an anisotropic pattern depending upon the incident polarization (32); however, a more detailed mechanism is still under investigation. Next, we observed azo-polymer deformation by rotating the polarization direction by 90 degrees, as indicated by the green arrow in Fig. 2(b) (see also Movie S2). As expected, the azo-polymer film formed a structure similar to that shown in Fig. 2(a), but the pattern was rotated by 90 degrees. In this case, the heights of the two lobes were almost the same, with better symmetry in the pattern.

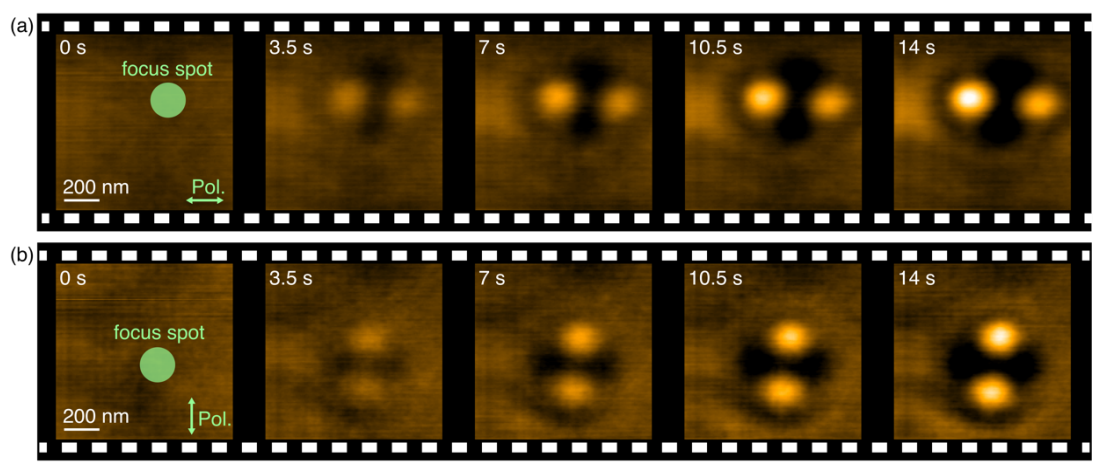

**Figure 2.** (a) HS-AFM images of azo-polymer film under the laser irradiation, obtained at 2 fps. (b) HS-AFM images of azo-polymer film under the laser irradiation with the polarization direction rotated by 90 degrees with respect to that in the case of (a). Green arrows indicate the polarization direction of the incident laser.

**Nanoscale analysis of the azo-polymer deformation process**

The topographic line profiles based on the HS-AFM movies allow detailed and quantitative analysis of the temporal changes in deformation with high spatiotemporal resolution. Figure 3(a) shows the HS-AFM image with an irradiation time of 14 s, which is identical to that of Fig. 2(a). The overlaid height line profiles from all frames along the white arrow in the image are shown in Fig. 3(b) (see also Movie S3). The line profile analysis quantitatively visualized the changes in the surface topography, showing the gradual formation of two lobes and a pit between them immediately after laser irradiation, where their motions are indicated by the arrows. The left lobe was twice as high as the right lobe. To investigate the changes in more detail, the heights of the two lobes and pit were plotted as a function of the laser irradiation duration, as shown in Fig. 3(c). The heights of the lobes increase linearly with estimated growth rates of 1.60 nm/s for the left lobe and 0.91 nm/s for the right lobe. They eventually increased up to 24.0 nm for the left lobe and 14.1 nm for the right lobe after 14 s of laser irradiation. The height at the pit position was almost the same or slightly increased in the first few frames, as also seen in the line profiles in Fig. 3(b). Subsequently, the height gradually and linearly decreased to approximately -4.5 nm after 14 s of laser exposure. We assume that this is because the two lobes were close to each other at the beginning, resulting in an increase in the height at the center. As the two lobes moved away from each other, the height of the center part decreased, leading to the formation of the pit. Figure 3(d) shows the temporal changes in the distance between the two lobes. Notably, this progression exhibited a nonlinear change; the lobes moved away rapidly in the beginning and decelerated subsequently. Given the diffraction-limited focal spot size of ~220 nm, the lobes moved faster when they were initially close to the focal spot. They slowed down as they were distanced from the focal spot. The final distance between the lobes was 395 nm after 14 s. We conducted the same analysis for the HS-AFM images shown in Fig. 2(b), in which the polarization direction was rotated by 90 degrees. Figure 3(e) shows the HS-AFM image of the azo-polymer film, shown in Fig. 2(b) for the irradiation time of 14 s. The height line profiles along the white arrow in Fig. 3(e) are shown in Fig. 3(f) (see also Movie S4). In this case, the height line profiles of the two lobes were symmetrical. The upper and lower lobes grew at rates of 1.02 and 0.97 nm/s, respectively, as shown in Fig. 3(g). Their heights eventually increased up to 14.3 nm for the upper lobe and 14.1 nm for the lower lobe, respectively. With regard to the height change at the position of the pit, we observed a behavior similar to that shown in Fig. 3(c). As seen in Fig. 3(h), a similar nonlinear change was also observed in the distance between the lobes, although the final distance was 325 nm, which is slightly different from the case shown in Fig. 3(d). Thus, HS-AFM can accurately and quantitatively analyze how individual azo-polymer structures are formed in real time, which is difficult to achieve using conventional AFM. These results clearly demonstrate the usefulness of HS-AFM observations. It should also be emphasized that the two structures shown in Fig. 3(a) and Fig. 3(e) were created under the same experimental conditions, except for the polarization direction; however the resulting structures were slightly different. For example, the pattern in Fig. 3(e) was symmetric, whereas that in Fig. 3(a) was asymmetric. The distance between the two lobes also differed between the two cases. We assume that this difference is due to a slight difference in the focal position or optical alignment. Although we attempted to maintain the same experimental conditions as much as possible, even a very small difference in the beam profiles of the laser focus can drastically affect the deformation pattern. (23) Because a small difference can cause different structures, it is highly beneficial to observe the deformation process of individual structures using HS-AFM, rather than obtaining the averaged information of several different structures by conventional AFM.

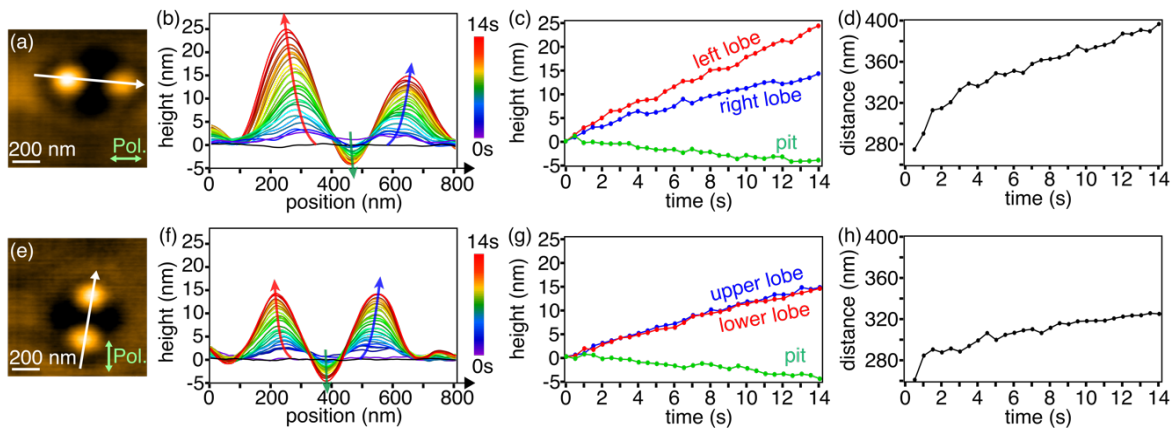

**Figure 3.** (a) HS-AFM image of the azo-polymer film, which is the same as the one shown in Fig. 2(a), for the irradiation time of 14 s. (b) Height line profiles obtained along the white arrow in (a), obtained for all frames of the HS-AFM movie at the same line position to investigate the temporal change. (c) Temporal height changes of the two lobes and pit, analyzed from (b). (d) Temporal change in the distance between the two lobes, analyzed from (b). (e) HS-AFM image same as the one shown in Fig. 2(b) for the irradiation time of 14 s. The incident polarization was rotated by 90° with respect to that in (a), as indicated by the green arrows. (f) Height line profiles obtained along the white arrow in (e), obtained for all frames of the HS-AFM movie. (g) Temporal height changes of the two lobes and the pit, analyzed from (f). (h) Temporal change in the distance between the two lobes, analyzed from (f).

## Azo-polymer deformation induced by z-polarization

In addition to the linear polarization parallel to the substrate, we observed the deformation process with *z*-polarization, which is a linear polarization perpendicular to the substrate. We used a *z*-polarizer (ZPol, Nanophoton) to generate *z*-polarization. Figure 4(a) shows the clipped HS-AFM images demonstrating the surface morphology change of the azo-polymer film under laser irradiation with *z*-polarization (see also Movie S5). Unlike horizontally polarized laser irradiation, a single lobe appeared at the center and a concentric circular pit surrounding the lobe was formed, which is consistent with a previous study. (25) Over time, the center lobe became higher, and the concentric circular pit became deeper. As shown in Fig. 4(b) (see also Movie S6), we quantitatively analyzed the deformation dynamics using the height line profile obtained along the white arrow in the image in Fig. 4(a). Here, pit1 and pit2 indicate the bottom and upper parts of the concentric pit, respectively, as indicated in Fig. 4(a). The time evolution of the lobe height is plotted in Fig. 4(c), showing a monotonic and linear increase at a rate of 0.63 nm/s. It eventually increased to 8.7 nm. The pit height decreased monotonically and linearly at a rate of -0.29 nm/s, finally reducing to -4.0 nm after 14 s at both the pit1 and pit2 regions.

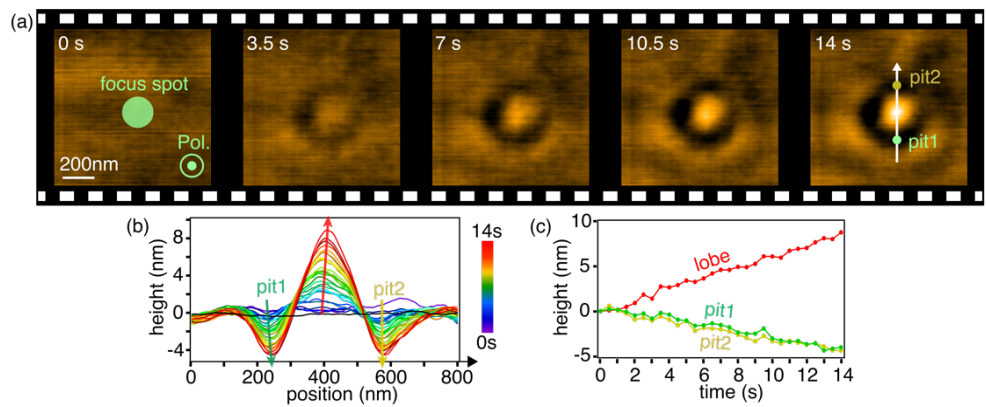

**Figure 4.** (a) HS-AFM images of the azo-polymer film under laser irradiation with the polarization perpendicular to the film (*z*-polarization). (b) Height line profile along the white arrow shown in (a) for all frames of the HS-AFM movie. (c) Temporal height changes of the lobe and pit.

**Influence of high-speed tip scanning on azo-polymer deformation**

We succeeded in observing the deformation process of the azo-polymer film by HS-AFM. However, even though HS-AFM has been recognized as a gentle technique, owing to the soft cantilever and sophisticated control system such as dynamic proportional-integral-differential (PID) feedback, such that even fragile proteins can be observed with negligible influence on their functions, high-speed scanning of the AFM tip may affect the deformation process of the azo-polymer and the resulting structure. Thus, we evaluated whether tip scanning affected the deformation process of the azo-polymer film or not. We induced azo-polymer deformation for 15 s using a *z*-polarized incident laser with observing the process by HS-AFM. In contrast, as a control experiment, deformation was induced for 15 s without HS-AFM observation. HS-AFM observation was performed after 15 s of laser irradiation, which means that high-speed tip scanning did not affect the deformation process. The same experiment was repeated three times for each condition. Figures 5(a) and (c) show HS-AFM images of the azo-polymer after deformation, with and without HS-AFM observations during the deformation process, respectively. Although all HS-AFM images showed slight differences because a small difference in the experimental conditions could affect the resulting structure, the overall surface morphologies were considerably similar. Figures 5(b) and (d) show the height line profiles along the arrows marked on each HS-AFM image. Although random differences were observed among the line profiles, they were considerably similar. The average height of the lobes was approximately 10 nm in both cases, regardless of whether HS-AFM observations were conducted during the deformation process. Therefore, we experimentally verified that HS-AFM is applicable to the observation of photoinduced movements of organic molecules such as azo-polymers.

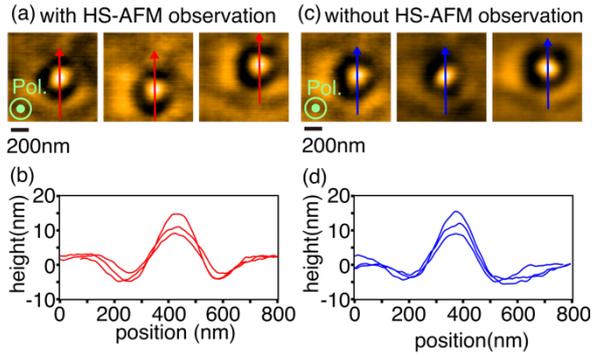

**Figure 5.** (a) HS-AFM images of the azo-polymer film after deformation, in which HS-AFM observations were conducted during deformation. (b) Height line profiles obtained along the lines marked in (a). (c) HS-AFM images of the azo-polymer film after deformation, in which HS-AFM observations were conducted after deformation. (d) Height line profiles obtained along the lines in (c).

## CONCLUSION AND DISCUSSION

In this study, we successfully demonstrated the *in-situ* real-time observation of the photoinduced surface deformation of an azo-polymer thin film using a tip-scan HS-AFM combined with an inverted optical microscope. The HS-AFM movie visualized the photo-induced changes in the surface morphology with high spatiotemporal resolution, showing the formation of unique patterns depending on the polarization direction. Moreover, the time evolution of the height line profile enabled precise and quantitative analysis. In addition, we found that a possible slight difference in the experimental conditions, such as optical alignment, significantly affected the resulting morphology. This fact suggests the importance of continuous observation of individual structures from the beginning to the end of deformation by HS-AFM, which is challenging with conventional AFM. We believe that *in-situ* HS-AFM observations will play a crucial role in further elucidating the mechanisms of azo-polymer deformation.

Although we have succeeded in observing the dynamic process of azo-polymer deformation, the imaging rate was still limited to 2 fps in this study. This is because the HS-AFM measurements had to be performed in air, as the current sample was not very resistant to water. In general, the quality factor of the cantilever in air is several hundreds or tens of times higher than that in solution, which slows the transient response of the cantilever and thus hinders high-speed imaging. By observing the deformation process in water, it would be possible to capture faster dynamic processes in detail under different conditions, such as high laser power causing rapid movements. More details of the deformation mechanism can be further understood through HS-AFM observations under various conditions, such as different laser wavelengths, polarization conditions, and types of azo-polymer molecules. For example, previous reports have shown the formation of unique spiral relief on azo-polymer films using optical vortex. (24,26) Furthermore, this technique would be useful not only for azo-polymers but also for many other samples in different fields. The potential of tip-scan HS-AFM has been extended to various fields by combining it with optical techniques. Our important demonstration of tip-scan HS-AFM would significantly stimulate researchers in diverse research fields, contributing to novel discoveries in the future in research fields from biology to many others including material science and physical chemistry.

**Author contributions**

Takayuki Umakoshi conceived and designed this project. K.Y. performed all experiments, analyzed the results, and wrote the manuscript. K.Y., FY.C., H.W., S.Y., Takayuki Uchihashi, and Takayuki Umakoshi constructed the setup for experiments. H.I. prepared the sample. H.I. and Takayuki Umakoshi supervised this research. All authors contributed to the discussion and finalization of the manuscript.

**Conflicts of interest**

There are no conflicts to declare.

**Acknowledgements**

This work was supported in part by JSPS KAKENHI Grant Number 20H02658, JSPS KAKENHI Grant Number 23H04590, JST PRESTO Grant Number JPMJPR19G2, and Takahashi Industrial and Economic Research Foundation.

**References**

(1) Ando, T.; Kodera, N.; Takai, E.; Maruyama, D.; Saito, K.; Toda, A. A high-speed atomic force microscope for studying biological macromolecules. *Proc. Natl. Acad. Sci.* **2001**, *98* (22), 12468–12472. DOI: 10.1073/pnas.211400898.

(2) Kodera, N.; Yamamoto, D.; Ishikawa, R.; Ando, T. Video imaging of walking myosin V by high-speed atomic force microscopy. *Nature* **2010**, *468* (7320), 72–76. DOI: 10.1038/nature09450.

(3) Uchihashi, T.; Iino, R.; Ando, T.; Noji, H. High-Speed Atomic Force Microscopy Reveals Rotary Catalysis of Rotorless F1-ATPase. *Science* **2011**, *333* (6043), 755–758. DOI: 10.1126/science.1205510.

(4) Shibata, M.; Nishimasu, H.; Kodera, N.; Hirano, S.; Ando, T.; Uchihashi, T.; Nureki, O. Real-space and real-time dynamics of CRISPR-Cas9 visualized by high-speed atomic force microscopy. *Nat. Commun.* **2017**, *8* (1), 1430. DOI: 10.1038/s41467-017-01466-8.

(5) Kobayashi, M.; Sumitomo, K.; Torimitsu, K. Real-time imaging of DNA-streptavidin complex formation in solution using a high-speed atomic force microscope. *Ultramicroscopy* **2007**, *107* (2–3), 184–190. DOI: 10.1016/j.ultramic.2006.07.008.

(6) Shibata, M.; Yamashita, H.; Uchihashi, T.; Kandori, H.; Ando, T. High-speed atomic force microscopy shows dynamic molecular processes in photoactivated bacteriorhodopsin. *Nat. Nanotechnol.* **2010**, *5* (3), 208–212. DOI: 10.1038/nnano.2010.7.

(7) Casuso, I.; Khao, J.; Chami, M.; Paul-Gilloteaux, P.; Husain, M.; Duneau, J.-P.; Stahlberg, H.; Sturgis, J. N.; Scheuring, S. Characterization of the motion of membrane proteins using high-speed atomic force microscopy. *Nat. Nanotechnol.* **2012**, *7* (8), 525–529. DOI: 10.1038/nnano.2012.109.

(8) Watanabe-Nakayama, T.; Ono, K.; Itami, M.; Takahashi, R.; Teplow, D. B.; Yamada, M. High-speed atomic force microscopy reveals structural dynamics of amyloid ß$_{1\text{-}42}$ aggregates. *Proc. Natl. Acad. Sci. U. S. A.* **2016**, *113* (21), 5835–5840. DOI: 10.1073/pnas.1524807113.

(9) Uchihashi, T.; Watanabe, Y.; Nakazaki, Y.; Yamasaki, T.; Watanabe, H.; Maruno, T.; Ishii, K.; Uchiyama, S.; Song, C.; Murata, K.; Iino, R.; Ando, T. Dynamic structural states of ClpB involved in its disaggregation function. *Nat. Commun.* **2018**, *9* (1), 2147. DOI: 10.1038/s41467-018-04587-w.

(10) Umakoshi, T.; Udaka, H.; Uchihashi, T.; Ando, T.; Suzuki, M.; Fukuda, T. Quantum-dot antibody conjugation visualized at the single-molecule scale with high-speed atomic force microscopy. *Colloids Surf. B* **2018**, *167*, 267–274. DOI: 10.1016/j.colsurfb.2018.04.015.

(11) Nishiguchi, S.; Furuta, T.; Uchihashi, T. Multiple dimeric structures and strand-swap dimerization of E-cadherin in solution visualized by high-speed atomic force microscopy. *Proc. Natl. Acad. Sci. U. S. A.* **2022**, *119* (30), e2208067119. DOI: 10.1073/pnas.2208067119.

(12) Ganser, C.; Uchihashi, T. Microtubule self-healing and defect creation investigated by in-line force measurements during high-speed atomic force microscopy imaging. *Nanoscale* **2019**, *11* (1), 125–135. DOI: 10.1039/C8NR07392A.

(13) Fukuda, S.; Ando, T. Faster high-speed atomic force microscopy for imaging of biomolecular processes. *Rev. Sci. Instrum.* **2021**, *92* (3), 033705. DOI: 10.1063/5.0032948.

(14) Marchesi, A.; Umeda, K.; Komekawa, T.; Matsubara, T.; Flechsig, H.; Ando, T.; Watanabe, S.; Kodera, N.;

Franz, C. M. An ultra-wide scanner for large-area high-speed atomic force microscopy with megapixel resolution. *Sci. Rep.* **2021**, *11* (1), 13003. DOI: 10.1038/s41598-021-92365-y.

(15) Chan, F.-Y.; Kurosaki, R.; Ganser, C.; Takeda, T.; Uchihashi, T. Tip-scan high-speed atomic force microscopy with a uniaxial substrate stretching device for studying dynamics of biomolecules under mechanical stress. *Rev. Sci. Instrum.* **2022**, *93* (11), 113703. DOI: 10.1063/5.0111017.

(16) Suzuki, Y.; Sakai, N.; Yoshida, A.; Uekusa, Y.; Yagi, A.; Imaoka, Y.; Ito, S.; Karaki, K.; Takeyasu, K. High-speed atomic force microscopy combined with inverted optical microscopy for studying cellular events. *Sci. Rep.* **2013**, *3* (1), 2131. DOI: 10.1038/srep02131.

(17) Fukuda, S.; Uchihashi, T.; Iino, R.; Okazaki, Y.; Yoshida, M.; Igarashi, K.; Ando, T. High-speed atomic force microscope combined with single-molecule fluorescence microscope. *Rev. Sci. Instrum.* **2013**, *84* (7), 073706. DOI: 10.1063/1.4813280.

(18) Umakoshi, T.; Fukuda, S.; Iino, R.; Uchihashi, T.; Ando, T. High-speed near-field fluorescence microscopy combined with high-speed atomic force microscopy for biological studies. *Biochim. Biophys. Acta. Gen. Subj.* **2020**, *1864* (2), 129325. DOI: 10.1016/j.bbagen.2019.03.011.

(19) Kumar, G. S.; Neckers, D. C. Photochemistry of Azobenzene-Containing Polymers *Chem. Rev.* **1989**, 89, 1915-1925.

(20) Yager, K. G.; Barrett, C. J. All-optical patterning of azo polymer films. *Curr. Opin. Solid State Mat. Sci.* **2001**, *5* (6), 487–494. DOI: 10.1016/S1359-0286(02)00020-7

(21) Gilbert, Y.; Bachelot, R.; Royer, P.; Bouhelier, A.; Wiederrecht, G. P.; Novotny, L. Longitudinal anisotropy of the photoinduced molecular migration in azobenzene polymer films. *Opt. Lett.* **2006**, *31* (5), 613–615. DOI: 10.1364/OL.31.000613.

(22) Ishitobi, H.; Tanabe, M.; Sekkat, Z.; Kawata, S. The anisotropic nanomovement of azo-polymers. *Opt. Express* **2007**, *15* (2), 652–659. DOI: 10.1364/OE.15.000652.

(23) Ishitobi, H.; Shoji, S.; Hiramatsu, T.; Sun, H.-B.; Sekkat, Z.; Kawata, S. Two-photon induced polymer nanomovement. *Opt. Express* **2008**, *16* (18), 14106–14114. DOI: 10.1364/OE.16.014106.

(24) Ambrosio, A.; Marrucci, L.; Borbone, F.; Roviello, A.; Maddalena, P. Light-induced spiral mass transport in azo-polymer films under vortex-beam illumination. *Nat. Commun.* **2012**, *3* (1), 989. DOI: 10.1038/ncomms1996.

(25) Ishitobi, H.; Nakamura, I.; Kobayashi, T.; Hayazawa, N.; Sekkat, Z.; Kawata, S.; Inouye, Y. Nanomovement of Azo Polymers Induced by Longitudinal Fields. *ACS Photonics* **2014**, *1* (3), 190–197. DOI: 10.1021/ph400052b.

(26) Watabe, M.; Juman, G.; Miyamoto, K.; Omatsu, T. Light induced conch-shaped relief in an azo-polymer film. *Sci. Rep.* **2014**, *4* (1), 4281. DOI: 10.1038/srep04281.

(27) Natansohn, A.; Rochon, P.; Gosselin, J.; Xie, S. Azo polymers for reversible optical storage. 1. Poly[4'-[[2-(acryloyloxy)ethyl]ethylamino]-4-nitroazobenzene]. *Macromolecules* **1992**, *25* (8), 2268–2273. DOI: 10.1021/ma00034a031.

(28) Tanchak, O. M.; Barrett, C. J. Light-Induced Reversible Volume Changes in Thin Films of Azo Polymers: The Photomechanical Effect. *Macromolecules* **2005**, *38* (25), 10566–10570. DOI:

10.1021/ma051564w.

(29) Zhao, Y.; Lu, Q.; Li, M.; Li, X. Anisotropic Wetting Characteristics on Submicrometer-Scale Periodic Grooved Surface. *Langmuir* **2007**, *23* (11), 6212–6217. DOI: 10.1021/la0702077.

(30) Sekkat, Z.; Knoll, W. Creation of second-order nonlinear optical effects by photoisomerization of polar azo dyes in polymeric films: theoretical study of steady-state and transient properties. *J. Opt. Soc. Am. B* **1995**, *12* (10), 1855–1867. DOI: 10.1364/JOSAB.12.001855.

(31) Karageorgiev, P.; Neher, D.; Schulz, B.; Stiller, B.; Pietsch, U.; Giersig, M.; Brehmer, L. From anisotropic photo-fluidity towards nanomanipulation in the optical near-field. *Nat. Mater.* **2005**, *4* (9), 699–703. DOI: 10.1038/nmat1459.

(32) Lee, S.; Kang, H. S.; Park, J.-K. Directional Photofluidization Lithography: Micro/Nanostructural Evolution by Photofluidic Motions of Azobenzene Materials. *Adv. Mater.* **2012**, *24* (16), 2069–2103. DOI: 10.1002/adma.201104826.

# *Supplementary Information*

In-situ real-time observation of photo-induced nanoscale azo-polymer motions using high-speed atomic force microscopy combined with an inverted optical microscope

Keishi Yang[a], Feng-Yueh Chan[b], Hiroki Watanabe[c], Shingo Yoshioka[a], Yasushi Inouye[ad], Takayuki Uchihashi[bc], Hidekazu Ishitobi[ad], Prabhat Verma[a], and Takayuki Umakoshi*[ae]

a Department of Applied Physics, Osaka University, 2-1 Yamadaoka, Suita, Osaka 565-0871, Japan

b Department of Physics, Nagoya University, Aichi 464-8602, Japan

c Exploratory Research Center on Life and Living Systems (ExCELLS), National Institutes of Natural Sciences, Higashiyama 5-1, Myodaiji, Okazaki, 444–0864 Japan

d Graduate School of Frontier Biosciences, Osaka University, 1-3 Yamadaoka, Suita, Osaka 565-0871, Japan

e Institute of Advanced Co-Creation Studies, Osaka University, 2-1 Yamadaoka, Suita, Osaka 565-0871, Japan

*E-mail: umakoshi@ap.eng.osaka-u.ac.jp

**Movie S1**: https://youtu.be/QVU2XmQgJIQ
HS-AFM movie of the azo-polymer deformation process. Linearly polarized light was used. The polarization direction and focus spot are indicated by the green arrow and circle, respectively, in Fig. 3(a). The movie was captured in a 1 μm$^2$ range, 150 × 150 pixels, at 2 fps.

**Movie S2**: https://youtu.be/QHGA16JOZb4
HS-AFM movie of the azo-polymer deformation process. Linearly polarized light was used. The polarization direction and focus spot are indicated by the green arrow and circle in Fig. 3(b). The movie was captured in a 1 μm$^2$ range, 150 × 150 pixels, at 2 fps.

**Movie S3**: https://youtu.be/4njzSaESFEE
Time evolution of the height line profiles shown in Fig. 4(b).

**Movie S4**: https://youtu.be/mg8P2PkQ9V8
Time evolution of the height line profiles shown in Fig. 4(f).

**Movie S5**: https://youtu.be/az_BtSURsqg

HS-AFM movie of the azo-polymer deformation process. Z-polarized light was used. The focus spot position is indicated by the green mark in Fig. 5(a). The movie was captured in a 1 μm$^2$ range, 150 × 150 pixels, at 2 fps.

**Movie S6**: https://youtu.be/EDNyLpQFATM

Time evolution of the height line profiles shown in Fig. 4(b).